\documentclass[11pt,a4paper]{article}
\usepackage{times,latexsym}
\usepackage{url}
\usepackage[T1]{fontenc}

\usepackage{tikz}
\definecolor{color003e72}{RGB}{0, 62, 144}
\definecolor{Green}{rgb}{0.0, 0.5, 0.0}
\usepackage{pgfplots}
\pgfplotsset{compat=1.6}
\usepackage{caption}
\usepackage{subcaption} % for subfigure
\usepackage{graphicx}
\usepackage{multirow}
\usepackage{tabularx}
\usepackage{colortbl} % row color
\usepackage{threeparttable} % table footnote
\usepackage{booktabs}       % professional-quality tables
\usepackage{balance} % balance refs
\usepackage{float}
\usepackage{hyperref}

\usepackage{pifont}
\usepackage{amsmath,amsfonts,bm}
\usepackage[bb=dsserif]{mathalpha}

\def\eqref#1{equation~\ref{#1}}
\def\Eqref#1{Equation~\ref{#1}}
\def\1{\bm{1}}

\DeclareMathAlphabet{\mathsfit}{\encodingdefault}{\sfdefault}{m}{sl}
\SetMathAlphabet{\mathsfit}{bold}{\encodingdefault}{\sfdefault}{bx}{n}

\newcommand{\KL}{\mathrm{KL}}

\DeclareMathOperator*{\argmin}{\arg\!\min}

\newcommand{\ind}{\perp\!\!\!\!\perp}

\usepackage[acceptedWithA]{tacl2021v1}
\usepackage{xspace,mfirstuc,tabulary}

\newif\iftaclinstructions
\taclinstructionsfalse % AUTHORS: do NOT set this to true
\iftaclinstructions
\renewcommand{\confidential}{}
\renewcommand{\anonsubtext}{(No author info supplied here, for consistency with
TACL-submission anonymization requirements)}
\newcommand{\instr}
\fi

\iftaclpubformat % this "if" is set by the choice of options

\else

\fi

\title{Learning Speech Representations with Variational Predictive Coding}

\author{
  Sung-Lin Yeh
  \\
  \texttt{\small sunglin.yeh@ed.ac.uk}
  \And
  Peter Bell
  \\
  \texttt{\small Peter.Bell@ed.ac.uk}
  \And
  Hao Tang
  \\
  \texttt{\small hao.tang@ed.ac.uk}
}
\author{{Sung-Lin Yeh \qquad Peter Bell \qquad Hao Tang} \\              
Institute for Language, Cognition and Computation \\
School of Informatics, University of Edinburgh \\
10 Crichton Street, Edinburgh EH8 9AB\\
\texttt{\small sunglin.yeh@ed.ac.uk}\quad \texttt{\small Peter.Bell@ed.ac.uk} \quad \texttt{\small hao.tang@ed.ac.uk}}

\date{}

\begin{document}
\maketitle
\begin{abstract}
%
% Current problems of the abstract:
%   1) again not all SSL approaches can be
%      cast as predictive coding
%   2) the way VQ is introduced is awkward
%   3) predictive coding is not even
%      introduced
%
% > The second sentence? Also the first sentence: 
%   the idea of training models to predict different parts of the input.
%
% Maybe we should just be open and say we
% are only propose a framework that
% can cover a large family of training
% objectives (but not all).
%
% > I only mention the generalization to `generative' approaches, not all ssl approaches.
%
% The claimed contributions are:
%   1) generality
%   2) clear separation of objective and
%      optimiation
%   3) improving HuBERT
%   4) controlled comparison
%
% > what about the vairational perspective? (2) should be the benefit of (1)?
%

% While several approaches have attempted to extend HuBERT, one of the best known speech representation learning
% objectives, but the underlying principles of the HuBERT objective has yet been explored.
Despite being the best known objective for learning speech representations, the HuBERT objective has not been further developed and improved.
We argue that it is the lack of an underlying principle that stalls the development,
and, in this paper, we show that predictive coding under a variational view is the principle behind the HuBERT objective.
Due to its generality, our formulation provides opportunities to improve parameterization and optimization, and we show two simple modifications that bring immediate improvements to the HuBERT objective.
% We demonstrate that methods like HuBERT and VQ-APC can be viewed as specific instances 
% of our general variational framework, allowing deeper insight into their design and optimization.
% By addressing deficiencies the optimization of HuBERT objective, such as independent optimization of HuBERT components, 
% we propose joint optimization strategies that lead to better optimization.
In addition, the predictive coding formulation has tight connections to various other objectives, such as APC, CPC, wav2vec, and BEST-RQ.
Empirically, the improvement in pre-training brings significant improvements to four downstream tasks: phone classification, f0 tracking, speaker recognition, and automatic speech recognition, highlighting the importance of the predictive coding interpretation.
% We validate the framework under different pretraining configurations and diverse downstream tasks, including other pre-training approaches
% it generalizes or relates to.
\end{abstract}

\section{Introduction}

Self-supervised learning has been the most successful approach to semi-supervised learning, leveraging unlabeled data for various downstream tasks \citep{zhang2022bigssl}.
The impact of self-supervised learning in speech processing has now been extended to the pursuit of more accessible yet compact, discrete representations to interact with language models \citep{lakhotia2021generative,borsos2023audiolm,wang2023neural}.
Behind these successes, other than the ease of scaling with Transformers, the training objectives are the major factor that makes the advancement possible.

The training objective of HuBERT \citep{hsu2021hubert} is undoubtedly the most successful self-supervised objective.
The HuBERT objective combines quantization and masked prediction as its main components.
However, those choices are supported by empirical evidence rather than an underlying principle, and it becomes difficult to further improve the HuBERT objective and to design novel objectives.
This is evidenced by the fact that subsequent work has focused on data augmentation (e.g., WavLM \citep{chen2022wavlm}), simplification of training (e.g., BEST-RQ \citep{chiu2022self} and MelHuBERT \citep{lin2022melhubert}), 
pairing with other objectives (e.g., w2v-BERT \citep{chung2021w2v}, DinoSR \citep{liu2024dinosr}, 
and MS-HuBERT \citep{yadav2024ms}), 
training with different resolution \citep{shi2023multi}; none has improved the HuBERT objective itself.

In this paper, we will show that predictive coding is the underlying principle of HuBERT.
It is not difficult to see the lineage of predictive coding in HuBERT.
HuBERT is inspired by wav2vec 2.0 \citep{baevski2020wav2vec} and DeepCluster \citep{caron2018deep}, 
and wav2vec 2.0 is in turn influenced by BERT \citep{devlin2018bert} and contrastive predictive coding (CPC) \citep{oord2018representation}.
However, the HuBERT objective looks distinctively different from, say, the formulation in \citet{atal1971speech}, \citet{srinivasan1982predictive}, and \citet{rao1999predictive}.
What is missing is a general framework for predictive coding (to subsume masked prediction) and a discrete intermediate representation (to subsume quantization).

In this paper, the general framework we develop is a variational view of predictive coding and we present HuBERT as a special case.
We will derive a training objective from first principles, and will show how it relates, not only to the HuBERT objective but also to various other objectives, such as VQ-APC \citep{chung2020vector}, VQ-CPC \citep{niekerk2020vector}, wav2vec 2.0, and BEST-RQ.
Moreover, our derivation can spawn new objectives, and we will give an example that brings immediate improvement over HuBERT.
Once we have an objective, the optimization of it, i.e., the algorithm of learning representations, naturally decouples, and provides opportunities for improvement.
We will give an example of how optimizing the HuBERT objective with a different algorithm brings immediate improvement over HuBERT.
We will also validate how much the improvement on the self-supervised objective transfers to downstream performance.

\section{A Variational Framework for Predictive Coding}
\label{sec:predcoding}

To see how predictive coding relates to HuBERT,
in this section, we briefly review predictive coding,
and its variational formulation.

Given its long history, predictive coding comes in various forms.
Predictive coding at the conceptual level, as described in \citet{elias1955predictive}, involves an encoder on one end and a decoder on the other.
The encoder processes a signal as it comes in (e.g., frames of a speech utterance in streaming mode), predicts what comes next, and computes the residuals (or how off the prediction is).
The decoder receives the residuals, makes a prediction, and combines the two to reconstruct the signal.
The hope is that the residuals require fewer bits to send than the original signal, achieving the goal of coding.

When predictive coding later evolves into algorithms, the goal becomes predicting one part of a signal given the other.
For example, it is predicting the next wave sample given the past samples in \citet{atal1971speech}, and predicting the center pixel given the neighboring pixels in \citet{srinivasan1982predictive}.
A detailed exposition is beyond the scope of this paper and can be found in \citet{makhoul1975linear}, \citet{huang2011predictive}, and \citet{spratling2017review}.

The general idea of predicting one part of a signal given the other can
be formalized as follows.  Let $x$ be a signal, for example, wave samples
of a speech utterance.  Predictive coding is about learning $p(x_B |
x_A)$, or minimizing $-\log p(x_B | x_A)$, where $x_A$ and $x_B$ forms
a partition of $x$.  To learn the entirety of $x$, the partition is not
fixed but drawn stochastically, resulting in the objective
\begin{align}
\mathbb{E}_{(x_A, x_B) \sim \mathcal{M}(x)}[-\log p(x_B | x_A)],
\label{eq:pc}
\end{align}
where $\mathcal{M}(x)$ is a distribution over many partitions of $x$.
The formulation in \citet{atal1971speech} can be seen as choosing an
$\mathcal{M}$ to partition wave samples in the future and the past,
while in \citet{srinivasan1982predictive}, $\mathcal{M}$ is chosen
to partition the center pixel and the neighboring pixels.  For the
interest of this paper, the signal $x$ is a sequence of acoustic frames $x_1,
\dots, x_T$, and we partition the signal into
$x_A=\{x_i\}_{i \in A}$ and $x_B = \{x_i\}_{i \in B}$
based on two sets of time indices $A \subset \{1,\dots,T\}$
and $B = \{1,\dots,T\} \setminus A$.
It is not difficult to see that $x_A$ will be the masked frames
and $x_B$ will be the unmasked frames in HuBERT,
and we willl make this explicit in the next section.

From the coding perspective,
a few important components are missing in equation \ref{eq:pc}: the encoder, the decoder, and the message sent from the encoder to the decoder.
Suppose the encoder encodes $x_A$ into a message $z$ and sends $z$ to the decoder to infer $x_B$.
We assume that knowing $z$ is sufficient to infer $x_B$, i.e., $x_B \ind x_A \mid z$.
Otherwise, the compression is deemed lossy.
A variational upper bound of equation \ref{eq:pc} can then be written as
\begin{align}
  \mathbb{E}_{(x_A, x_B) \sim \mathcal{M}(x)} \Big[ \KL & \big[ q(z|x_B)\|p(z|x_A)\big] \label{eq:vlb} \\
  & - \mathbb{E}_{z \sim q(z|x_B)}[\log p(x_B | z)] \Big], \notag
\end{align}
where $q(z|x_B)$ is an auxiliary distribution of our choice.\footnote{See Appendix \ref{appendix:proof} for the derivation of our objective based on the variational lower bound.} 
The second term $\mathbb{E}_{z \sim q(z|x_B)}[\log p(x_B | z)]$ 
is known as the reconstruction (or distortion in coding),
where $p(z|x_A)$ is thought of as the encoder, $p(x_B | z)$ the decoder, and $z$ the message.
The variational formulation has the advantage of making
the encoder, decoder, and message explicit in the objective.
Equation \ref{eq:vlb} is known as the negative free energy or the negative evidence lower bound (negative ELBO), 
and this view of predictive coding is first made explicit in \citet{friston2009predictive} and later generalized in \citet{feldman2010attention}.
Our treatment adheres more to the variational lower bound in \citet{Kingma2014} and \citet{sohn2015learning} 
with the additional assumption that $x_B \ind x_A \mid z$.

\section{HuBERT as Predictive Coding}
\label{sec:hubert}

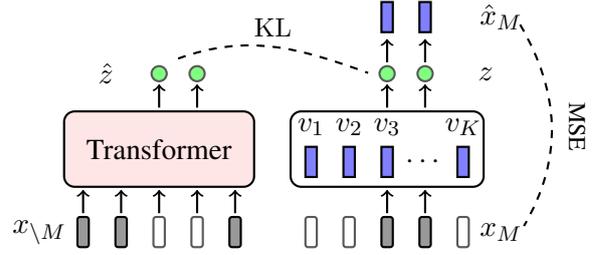
\begin{figure}
\centering
\begin{tikzpicture}[code/.style={draw, rectangle, fill=blue!50, inner sep=0.05cm, 
                                        minimum width=0.15cm, minimum height=0.4cm, thick}]

\node[code,rounded corners=1,fill=black!0,draw=black!65] (xb0) at (3.5, 0) {};
\node[code,rounded corners=1,fill=black!0,draw=black!65] (xb1) at (4, 0) {};
\node[code,rounded corners=1,fill=black!40] (xb2) at (4.5, 0) {};
\node[code,rounded corners=1,fill=black!40] (xb3) at (5, 0) {};
\node[code,rounded corners=1,fill=black!0,draw=black!65,label={right:$x_M$}] (xb4) at (5.5, 0) {};

\node[code,rounded corners=0] (xb2) at (4.5, 2.2+1.1-0.45) {};
\node[code,rounded corners=0] (xb3) at (5, 2.2+1.1-0.45) {};
\node[code,rounded corners=1,fill=black!0,draw=black!0,label={right:$\hat{x}_M$}] (xb4) at (5.5, 2.2+1.1-0.45) {};

\node[code,rounded corners=1,fill=black!40,label={left:$x_{\backslash M}$}] (xa0) at (0.5, 0) {};
\node[code,rounded corners=1,fill=black!40] (xa1) at (1, 0) {};
\node[code,rounded corners=1,fill=black!0,draw=black!65] (xa2) at (1.5, 0) {};
\node[code,rounded corners=1,fill=black!0,draw=black!65] (xa3) at (2, 0) {};
\node[code,rounded corners=1,fill=black!40] (xa4) at (2.5, 0) {};
\node [draw, rectangle, fill=red!10, thick, 
        rounded corners, minimum width=2.5cm, minimum height=1cm] (trans) at (1.5, 1.1) {Transformer};
\node[circle, thick, inner sep=0pt, minimum size=0.2cm, fill=green!50,draw=black!65] (za2) at (1.5, 2.1) {};
\node[circle, thick, inner sep=0pt, minimum size=0.2cm, fill=green!50,draw=black!65] (xa3) at (2, 2.1) {};
\node (z) at (0.8, 2.1) {$\hat{z}$};

\node[circle, thick, inner sep=0pt, minimum size=0.2cm, fill=green!50,draw=black!65] (xb2) at (4.5, 2.1) {};
\node[circle, thick, inner sep=0pt, minimum size=0.2cm, fill=green!50,draw=black!65] (xb3) at (5, 2.1) {};
\node (z) at (5.8, 2.1) {$z$};

\draw[->,thick] (0.5, 0.25) -- (0.5, 0.56);
\draw[->,thick] (1, 0.25) -- (1, 0.56);
\draw[->,thick] (1.5, 0.25) -- (1.5, 0.56);
\draw[->,thick] (2, 0.25) -- (2, 0.56);
\draw[->,thick] (2.5, 0.25) -- (2.5, 0.56);
\draw[->,thick] (2, 0.25+1.4) -- (2, 0.56+1.4);
\draw[->,thick] (1.5, 0.25+1.4) -- (1.5, 0.56+1.4);
\draw[->,thick] (4.5, 0.25+1.4) -- (4.5, 0.56+1.4);
\draw[->,thick] (5, 0.25+1.4) -- (5, 0.56+1.4);
\draw[->,thick] (4.5, 0.25+1.4+0.6) -- (4.5, 0.56+1.4+0.6);
\draw[->,thick] (5, 0.25+1.4+0.6) -- (5, 0.56+1.4+0.6);
\draw[->,thick] (4.5, 0.25) -- (4.5, 0.56);
\draw[->,thick] (5, 0.25) -- (5, 0.56);

\draw[dashed, thick] (1.75, 2.3) to [out=20, in=160] node[above] {\small $\KL$} (4.25, 0.56+1.55);
\draw[dashed, thick] (6.25, 2.2+1.1-0.55) to [out=-60, in=60] node[right, rotate=-90] {} (6.3, 0.08); 
\node[rotate=-90] (mse) at (7, 1.4) {\footnotesize MSE};

\node[code, label={above:$v_1$}] (v1) at (3.5, 0.93) {};
\node[code, label={above:$v_2$}] (v2) at (4, 0.93) {};
\node[code, label={above:$v_3$}] (v3) at (4.5, 0.93) {};
\node (v4) at (5, 0.93) {$\cdots$};
\node[code, label={above:$v_K$}] (v5) at (5.5, 0.93) {};
\draw[rounded corners, thick] (4.5-1.27, 0.6) rectangle (4.5+1.27, 1.6);
\end{tikzpicture}
\caption{HuBERT as variational predictive coding.
  The set $v_1, \dots, v_K$ are the codewords in the codebook,
  $x_{\backslash M}$ are the unmasked frames,
  and $x_M$ are the masked frames.
  There are two loss functions involved, the Kullback-Leibler divergence ($\KL$)
  and the mean-squared error (MSE).}
\label{fig:feedforward}
\end{figure}

Given the variational framework of predictive coding, we now turn to the HuBERT objective and discuss how it relates to predictive coding.
Recall that HuBERT training consists of two steps:
first quantizing the acoustic frames and second training a
Transformer to predict the cluster IDs of the quantized frames.
The HuBERT objective often refers to the cross entropy of predicting
the cluster IDs of each frame (shown as KL in Figure \ref{fig:feedforward}).
However, the cluster IDs of the quantized frames are produced by $k$-means,
so there is an implicit $\ell_2$ loss that measures
the distortion (or reconstruction) of $k$-means
(shown as MSE in Figure \ref{fig:feedforward}).

For the following subsections, we will detail how
the partition $\mathcal{M}$ and the parameterization of $q(z|x_A)$,
$p(x_B | z)$, and $p(z|x_A)$ are chosen, such that
the variational objective in Equation \ref{eq:vlb}
covers the cross entropy and the $\ell_2$ loss.
In particular, we will assume the latent variables are discrete
and correspond to the cluster IDs of frames.

\subsection{Masked Prediction}
\label{methods:mlm}

When training HuBERT, a mask is generated at random for every utterance, and frames in an utterance are partitioned into those that are masked and those that are not.
The objective is to predict the cluster IDs of the masked frames given the unmasked frames.
Formally, a mask is a subset of indices $M \subset \{1,\dots,T\}$,
and forms a partition $x_M = \{x_i\}_{i \in M}$ and $x_{\setminus M} = \{x_i\}_{i \not\in M}$.
Let $\mathcal{M}(x)$ be the stochastic process of generating masks for the utterance $x$, where typically a frame has a small probability to be the start of a span of frames being masked \citep{baevski2020wav2vec}.
With this choice of $\mathcal{M}$, the predictive coding objective (equation \ref{eq:vlb}) becomes
\begin{align}
\mathcal{L}_{\text{Masked}} = \mathbb{E}_{(x_{\setminus M}, x_M) \sim \mathcal{M}(x)} [\mathcal{L}_{x_{\setminus M}, x_M}]
\end{align}
where
\begin{align}
\mathcal{L}_{x_{\setminus M}, x_M} = \KL & \big[q(z|x_M)\|p(z|x_{\setminus M})\big] \notag \\
  & + \mathbb{E}_{z \sim q}[-\log p(x_M | z)].
\end{align}
Since the HuBERT objective is frame-wise, we assume $q(z|x_M)$ and
$p(x_M | z)$ factorize frame-wise, i.e., $q(z|x_M) = \prod_{i \in M} q(z_i | x_i)$
and $p(x_M | z) = \prod_{i \in M} p(x_i | z_i)$,
where $z_1, \dots, z_T$ are discrete latent variables for frames $x_1, \dots, x_T$.
After including the frame-wise independence, we have
\begin{align}
\mathcal{L}_{x_{\setminus M}, x_M}
  = \sum_{i \in M} & \mathbb{E}_{z_i \sim q} \big[\log q(z_i|x_i) \label{eq:vlb-mlm} \\
   - &\log p(z_i|x_{\setminus M}) - \log p(x_i | z_i)] \big]. \notag
\end{align}
Each $z_i \in \{1, \dots, K\}$ corresponds to the cluster ID of $x_i$, where $K$ is the total number of clusters.
Note that the second term
$\mathbb{E}_{z_i \sim q}[-\log p(z_i | x_{\setminus M})]$
is the cross entropy, and the last term
$\mathbb{E}_{z_i \sim q}[-\log p(x_i | z_i)]$
is the reconstruction.
We will discuss how these two terms become the cross entropy
and the $\ell_2$ loss of $k$-means in the HuBERT objective.
We will also discuss why
the first term $\mathbb{E}_{z_i \sim q}[\log q(z_i | x_i)]$,
the negative entropy, does not appear in the HuBERT objective.

\subsection{Quantization and Reconstruction}

In HuBERT, cluster IDs of frames that later serve as targets
for prediction are produced by $k$-means.
The $k$-means algorithm finds the ID of the closest centroid
for each individual frame, where closeness is defined by the $\ell_2$ loss.
To realize the quantization with $k$-means in our predictive coding framework,
we assign a point mass to the minimum and let
\begin{align}
q(z_i|x_i) = \mathbb{1}_{z_i = \argmin_{k=1, \dots, K} \|x_i - v_k\|^2},
\label{eq:argmin}
\end{align}
where $v_k$ is the $k$-th column of a matrix $V$, a codebook consisting of the centroids as columns.
Note that $q$ in principle can depend on $x_M$, but for this particular choice,
$q$ only depends on $x_i$.

The $k$-means objective is to minimize the distortion (or reconstruction) measured by the $\ell_2$ loss.
Given the codebook~$V$, using the cluster $z_i$ to reconstruct $x_i$ with the centroid $v_{z_i}$ gives a distortion $\|x_i - v_{z_i}\|^2$.
Naturally, we choose to parameterize $p(x_i|z_i)$ with a Gaussian and let
\begin{align}
p(x_i|z_i) = \frac{1}{(2\pi)^{d/2}}\exp\left(-\frac{1}{2}\|x_{i} - v_{z_{i}}\|^2\right),
\label{eq:rec}
\end{align}
where $d$ is the dimension of an acoustic frame.
The log of $p(x_i | z_i)$ gives the $\ell_2$ loss, i.e., the mean-squared error (MSE).
It is now clear that the term $\mathbb{E}_{z_i \sim q}[-\log p(x_i | z_i)]$ in equation \ref{eq:vlb-mlm} involves both the quantization of a frame $x_i$ and reconstructing it with the closest centroid $v_{z_i}$.
The negative entropy term $\mathbb{E}_{z_i \sim q}[\log q(z_i|x_i)]$ becomes 0 due to $q$ being a point mass.

\subsection{Predicting Quantized Frames}

Because $z_1, \dots, z_T$ correspond to the targets for prediction, we parameterize $p(z_i|x_{\setminus M})$
as a softmax, i.e.,
\begin{align}
p(z_i|x_{\setminus M}) = 
  \frac{\exp\left(\mathrm{enc}(x_{\setminus M})_i^\top u_{z_{i}}\right)}
    {\sum_{k=1}^K \exp\left(\mathrm{enc}(x_{\setminus M})_i^\top u_k\right)},
\label{eq:prior}
\end{align}
where $\mathrm{enc}(\cdot)$ is an encoder (typically a Transformer encoder),
$\mathrm{enc}(x_{\setminus M})_i$ is the $i$-th frame of the encoder output after taking the unmasked frames $x_{\setminus M}$ as input, and $u_k$ is the $k$-th column of the final linear layer~$U$.
The choice of $q(z_i|x_i)$ paired with the $p(z_i | x_{\setminus M})$ above completes the cross entropy $\mathbb{E}_{z_i \sim q}[-\log p(z_i | x_{\setminus M})]$ in equation~\ref{eq:vlb-mlm} for predicting the cluster IDs of frames.

\subsection{Two-Step Optimization}
\label{sec:optimization}

We have now instantiated the HuBERT objective from predictive coding.
The cross entropy term $\mathbb{E}_{z_i \sim q}[-\log p(z_i | x_{\setminus M})]$ and the reconstruction term $\mathbb{E}_{z_i \sim q}[-\log p(x_i | z_i)]$ are both present in the objective (equation \ref{eq:vlb-mlm}).
In principle, both terms should be optimized together,
but HuBERT takes a two-step approach, first finding the codebook by optimizing
the reconstruction (running $k$-means)
and then finding the Transformer parameters by optimizing the cross entropy.
The two-step approach is reminiscent to the variational view of expectation maximization \citep{neal1998view,attias1999variational}.
However, the codebook in HuBERT is never updated again after the first $k$-means,
and is said to be optimized offline as opposed to jointly with the Transformer
parameters.
This provides an opportunity to improve the optimization of the HuBERT objective.

\section{Extensions}
\label{sec:extensions}

Given how HuBERT is a special case of predictive coding, in this section, we discuss several immediate extensions that are made possible by our framework.

\subsection{Softening the Point Mass}

Even though the two-step optimization in HuBERT likely leads to suboptimal solutions, it turns out to be difficult to optimize both the cross entropy and the reconstruction together.
The difficulty stems from $q$ being a point mass, which cannot be optimized with gradient descent.
Instead of exact minimization, we can use soft-min and let
\begin{align}
q(z_i|x_i) = \frac{\exp\left(-\|x_i - v_{z_i}\|^2 / \tau\right)}
  {\sum_{k=1}^{K} \exp\left(-\|x_i - v_k\|^2 / \tau\right)},
  \label{eq:softmax-q}
\end{align}
where $\tau$ is the temperature.
As $\tau \to 0$, the distribution collapses to minimization or a hard $k$-means assignment \citep{Kulis2011RevisitingKN}, where each $x_i$ is assigned to the code $v_{z_i}$ with the smallest squared Euclidean distance.
With this parameterization, it is now possible to optimize the entire objective in equation \ref{eq:vlb-mlm} with gradient descent.

Since $q$ is no longer a point mass, we now have an additional negative entropy term $\mathbb{E}_{z_i \sim q}[\log q(z_i | x_i)]$ to optimize in equation \ref{eq:vlb-mlm}.
As entropy is maximized when $q$ is uniform, this term, when being minimized with other terms, encourages a diverse set of codes to be used and serves as a regularizer.
This is reminiscent to the diversity loss in wav2vec 2.0, and we will discuss the differences in later sections.

\subsection{Approximating the Expectation with Sampling}

Since $p(x_i | z_i)$ is parameterized in a relatively simple form, computing the expectation $\mathbb{E}_{z_i \sim q}[-\log p(x_i | z_i)]$ by enumerating all values of $z_i$, i.e., exact marginalization, is feasible.
However, exact marginalization is not always feasible when $p(x_i | z_i)$ becomes expensive to compute.
An alternative is to approximate the expectation with sampling.
There are various approaches to optimizing a function that involves sampling, and the simplest solution is to use Gumbel softmax \citep{jang2016categorical}.
We will compare marginalization and Gumbel softmax in the experiments.
Note that, the expectation is optimized offline with $k$-means in HuBERT, which will also be included in the comparisons.

\subsection{Future Prediction}
\label{sec:futurepred}

Our framework of predictive coding can also instantiate future prediction given the past.\footnote{Next-token prediction in language modeling, e.g., is a form of future prediction.}
The partition of a signal is a simpler than the masked prediction in that choosing a time point partitions the signal into the past and the future.
Formally, let $\mathcal{M}(x)$ be the stochastic process of choosing a time point $t$ in a signal $x$.
The past $x_{< t}$ and the future $x_{\geq t}$ forms a partition of $x$, and our autoregressive objective becomes
\begin{align}
\mathcal{L}_{\text{Future}} = \mathbb{E}_{(x_{< t}, x_{\geq t}) \sim \mathcal{M}(x)} [\mathcal{L}_{x_{< t}, x_{\geq t}}]
\end{align}
where
\begin{align}
\mathcal{L}_{x_{< t}, x_{\geq t}}
  = \sum_{i = t}^{T} & \mathbb{E}_{z_i \sim q} \big[ \log q(z_i|x_i) \label{eq:vlb-ar} \\
  & - \log p(z_i|x_{< i}) - \log p(x_i | z_i)]\big]. \notag
\end{align}
The form of equation \ref{eq:vlb-ar} is nearly identical to masked prediction in equation \ref{eq:vlb-mlm}, except the term $p(z_i | x_{< i})$.\footnote{The derivation of future prediction can be found in Appendix \ref{appendix:proof}.}
In terms of parameterization, $p(z_i | x_{< i})$ is typically modeled with a unidirectional LSTM or a causal Transformer; the rest of the terms remain the same.
Note that only the suffix of a signal participates in the objective, and $\mathcal{M}$ needs to be carefully chosen to avoid putting too much weight on the suffixes.
In practice, instead of choosing a time point at random, all frames are predicted with an equal amount of times \citep{oord2018representation,chung2019unsupervised}.

Since speech is generally smooth in time, an additional assumption $z_i \ind x_{i-\kappa+1:i} \mid x_{\leq i-\kappa}$ is commonly made for a small $\kappa > 0$ (e.g., in \citet{oord2018representation} and \citet{chung2019unsupervised}).
In other words, once we know the past frames $x_{\leq i-\kappa}$
close enough to the current time point $i$,
knowing additional few frames $x_{i-\kappa+1:i}$
does not add much information to $z_i$. 
The term $p(z_i | x_{< i})$ becomes $p(z_i | x_{\leq i-\kappa})$ under this assumption.

\section{Connections to Prior Work}
\label{sec:related}

We have demonstrated in Section \ref{sec:hubert} how our framework
instantiates HuBERT.
Given the generality of our framework, we can also instantiate other objectives that are similar to other self-supervised objectives.
In this section, we will discuss what our framework can achieve and how it differs from prior work,
including approaches based on likelihood and contrastive learning.
We do not consider combined objectives in this section, such as
w2v-BERT \citep{chung2021w2v}.

\subsection{APC and VQ-APC}

We begin with other variants of predictive coding for speech representation that optimizes the likelihood
\eqref{eq:pc}, in which $x_A$ is $x_{<t}$, $x_B$ is $x_{\geq t}$.
First, autoregressive predictive coding \citep{chung2019unsupervised,yang2022autoregressive} (APC) is a generalization of linear predictive coding \citep{atal1971speech,saito1967theoretical} (LPC), 
where a model is trained to predict future frames given the past.
Its vector-quantized variant, VQ-APC \citep{chung2020vector}, 
includes a quantization layer in the neural network while optimizing APC.
Both APC and VQ-APC optimize the likelihood, while our framework is based on the variational bound in equation \ref{eq:vlb}.
APC and VQ-APC in their original form in \citet{chung2019unsupervised} and \citet{chung2020vector} do not have the latent variables explicitly stated.
Our framework makes the choice of latent variables explicit, and the proposed auxiliary distribution 
offers more flexibility to estimate the likelihood, especially when marginalization of 
latent variables is not tractable.

Similar to our approach, \citet{yang2022autoregressive} and \citet{yeh2022autoregressive} make the latent variables explcit, leading to various other interpretations of APC with respect to, e.g., mutual information and co-training \citep{mcallester2018information}.

\subsection{MPC and DeCoAR}

MPC \citep{jiang2019improving,zhang2021transformer} and DeCoAR \citep{9053176} are both generalizations of APC,
replacing future prediction with masked prediction.
DeCoAR 2.0 \citep{ling2020decoar} adds quantization similar to VQ-APC.
Both optimize the likelihood in equation \ref{eq:pc}.
The difference between them lies in how the frames are masked---MPC masks many small spans while DeCoAR masks a single large span.

\subsection{HuBERT, WavLM, and BEST-RQ}
\label{sec:hubert-variants}

We have shown that HuBERT is a special case of our framework.
Subsequent work, such as WavLM \citep{chen2022wavlm} and BEST-RQ \citep{chiu2022self}, uses the same HuBERT objective.
WavLM improves over HuBERT with data augmentation.
BEST-RQ avoids updating the codebook, but requires a careful initialization.

When training HuBERT in \citet{hsu2021hubert}, there are multiple iterations, each of which trains a Transformer encoder from scratch.
What differs for each iteration are the training targets.
In the first iteration, quantized MFCC is used as targets, while in the second iteration, quantized hidden vectors from layer 9 of the first iteration are used as targets.
Our framework can also instantiate the second iteration with a pre-trained Transformer encoder from the first iteration. 
We can simply let
\begin{align}
    q(z_i|h_i) = \frac{\exp\left(-\|h_i - v_{z_i}\|^2 / \tau\right)} 
  {\sum_{k=1}^{K} \exp\left(-\|h_i - v_k\|^2 / \tau\right)},
\label{eq:softmax-q-iter2}
\end{align}
where the $i$-th frame $h_i$ now comes from the intermediate layer (e.g., 6th layer) of
the encoder from the first iteration.

\subsection{CPC, VQ-CPC, and wav2vec 2.0}
\label{sec:wav2vec}

CPC \citep{oord2018representation}, its vector-quantized variants, VQ-CPC \citep{niekerk2020vector}, 
wav2vec \citep{schneider2019wav2vec}, and wav2vec 2.0 \citep{baevski2020wav2vec}, 
are all based on noise contrastive estimation (NCE) \citep{smith-eisner-2005-contrastive,pmlr-v9-gutmann10a}.
In this section, we show how wav2vec 2.0 can be instantiated with our framework.

We derive their contrastive objective in wav2vec 2.0 from the cross entropy in the variational objective
 $\mathbb{E}_{z_i \sim q}[-\log p(z_i | x_{\setminus M})]$.
Recall that wav2vec 2.0 has a CNN followed by VQ and a Transformer,
where the VQ produces targets for contrastive learning.
We first choose
\begin{align}
q(z_i | x_M) &= \delta_{z_i=g(x_M)} \label{eq:w2v2-q}\\
p(z_i | x_{\setminus M}) &= \frac{\exp(\mathrm{sim}(\mathrm{enc}(x_{\setminus M})_i, z_i))}{
  \int \exp(\mathrm{sim}(\mathrm{enc}(x_{\setminus M})_i, z)) dz},
\end{align}
where $g(\cdot)$ is the CNN with the VQ, $\mathrm{enc}(\cdot)$ is the Transformer,
$\delta$ being the Dirac delta function, and $\mathrm{sim}(x, y) = \cos(x, y)$.
The choice of $p(z_i | x_{\setminus M})$ is similar to that of \Eqref{eq:prior} in our framework,
except a similarity function and a continuous $z$.
The cross entropy becomes
\begin{align}
    \mathbb{E}_{z_i \sim q}&[-\log p(z_i | x_{\setminus M})] \notag\\
    =&\log\frac{\exp(\mathrm{sim}(\mathrm{enc}(x_{\setminus M})_i, z_i))}{
   \int\exp(\mathrm{sim}(\mathrm{enc}(x_{\setminus M})_i, z')) dz}.
\end{align}
Note that $z_i$ here is the codeword after quantization, i.e., a continuous vector \citep{baevski2020wav2vec}.
The denominator in the cross entropy is in general difficult
to compute due to the integral.\footnote{In fact,
even when $\mathrm{sim}(x, y) = x^\top y$, i.e.,
$p(z_i | x_{\setminus M})$ being a von-Mises--Fisher distribution,
the denominator of $p$ is still difficult to compute.}
Inspired by NCE, \citet{oord2018representation} and \citet{baevski2020wav2vec}
approximate the integral by summing over a set of negative samples $S_i$. 
The cross entropy becomes
\begin{align}
    \log \frac{\exp(\mathrm{sim}(\mathrm{enc}(x_{\setminus M})_i, z_i))}{
  \sum_{z' \in S_i} \exp(\mathrm{sim}(\mathrm{enc}(x_{\setminus M})_i, z'))}.
  \label{eq:w2v2}
\end{align}
which is the objective used in wav2vec 2.0.
Consequently, minimizing the contrastive loss is analogue to minimizing the cross entropy between 
the masked frames and unmasked frames after quantization.

Several design choices in wav2vec 2.0 are taken verbatim by HuBERT, but they have a few key differences.
Both use a softmax to parameterize $p(z_i | x_{\setminus M})$, but wav2vec 2.0 assumes $z_i$ to be a continuous vector, while HuBERT assumes $z_i$ to be discrete.
As a consequence, wav2vec 2.0 uses a softmax over the negative and positive samples, while HuBERT uses a softmax over the codewords in a codebook.
The codebook in wav2vec 2.0 is trained together with the rest of the model.
In other words, there is no reconstruction and we can simply parameterize
$p(x_i | z_i)$ as a uniform distribution. 
In contrast, the codebook in HuBERT is trained with an offline $k$-means.
Compared to our framework, we have the choice to optimize the codebook
offline or jointly with the Transformer.

There is a additional diversity loss in wav2vec 2.0 that
computes the entropy of codeword usage within a batch,
to encourage the use of individual codewords.
The diversity loss was not meant to be part of the main objective
\cite{baevski2020wav2vec}, but more of a regularizer that improves
training.
However, our instantiation on HuBERT naturally has an entropy term
per frame that serves a similar purpose as the diversity loss.

\begin{table*}
\centering
\scalebox{0.88}{
\begin{tabular}{llllll}
\toprule
\textbf{Model} & $p(z|x_A)$ & $q(z|x_B)$ & $p(x_B|z)$ & \textbf{Codebook} & $\mathbb{E}_{z \sim q(z|x_B)}[\cdot]$ \\
\midrule
\multicolumn{3}{l}{\textbf{\textsc{Masked Prediction}}} &\\[1pt]
HuBERT Obj (eq \ref{eq:vlb-mlm}) 
              & softmax (eq \ref{eq:prior})  & point-mass (eq \ref{eq:argmin})
              & Gaussian (eq \ref{eq:rec})   & offline
              & single point \\
Masked-VPC (eq \ref{eq:vlb-mlm})   
              & softmax (eq \ref{eq:prior})  & soft-min (eq \ref{eq:softmax-q})
              & Gaussian (eq \ref{eq:rec})   & joint opt.
              & Gumbel / marginal \\
Masked-NCE (eq \ref{eq:vlb-mlm}) 
              & softmax (eq \ref{eq:w2v2})   & point-mass (eq \ref{eq:w2v2-q})
              & uniform                      & joint opt.
              & single point \\[2pt]
\midrule
\multicolumn{3}{l}{\textbf{\textsc{Future Prediction}}} &\\[1pt]
Future-VPC (eq \ref{eq:vlb-ar})
              & softmax (eq \ref{eq:prior})  & soft-min (eq \ref{eq:softmax-q})
              & Gaussian (eq \ref{eq:rec})   & joint opt.
              & Gumbel / marginal \\
VQ-APC (eq \ref{eq:pc}) 
              & softmax (eq \ref{eq:prior})  & n/a  
              & Gaussian (eq \ref{eq:rec})   & joint opt.
              & n/a \\
\bottomrule
\end{tabular}
}
\caption{A summary of objectives that can instantiated from our framework.
Each loss is charaterized by how the probability distributions are parameterized, how the code is optimized, and how the expectation is computed.
When $q$ is a point-mass, the expectation becomes an assignment to the point (single point).
The codebook can be either optimized with $k$-means (offline) or jointly optimized (online) with the Transformer.}
\label{tab:model-summary}
\end{table*}

\section{Experiments}
\label{sec:exp}

We have shown that our framework subsumes the HuBERT objective and the two are numerically identical.
In the experiments, we will study how the extensions in Section \ref{sec:extensions} can lead to immediate improvements over HuBERT.

\subsection{Experimental Settings}
\label{exp:setup}

Unless otherwise stated, models are built with a Transformer encoder, a linear projection after the encoder, and a codebook (consisting of the centroids in $k$-means).
We follow \citet{chung2019unsupervised}, \citet{jiang2019improving}, \citet{ling2020decoar}, \citet{misra21_interspeech}, \citet{chiu2022self}, and \citet{lin2022melhubert}, using Mel spectrograms as acoustic frames.
We extract 40-dimensional log Mel features with a 25 ms window and a 10 ms hop.
We concatenate every two contiguous frames, having 80-dimensional features with a 20 ms frame rate.
For masked prediction, we use a masking span of 4 frames (80 ms) with every frame being the start of a span with a probability of 0.2; spans may overlap.
For future prediction, we use a time shift (the $\kappa$ in Section \ref{sec:futurepred}) of 2 frames (40 ms).

In the experiments, we consider a \textsc{Base} setting similar to that
described by \citet{baevski2020wav2vec,hsu2021hubert}, using 12-layer Transformers
with 6 attention heads instead of 8.
We set the codebook size to 100 for all models, following the cluster size
in \citet{hsu2021hubert}.
We train models on the 360-hour subset of LibriSpeech \citep{panayotov2015librispeech}
for 150 epochs.
We detail hyperparameters in Appendix \ref{appendix:recipe}.

\subsection{Model Variants and Feature Summary}

Before comparing different optimization strategies and downstream performance, we summarize 
model variants evaluated in our experiments in Table~\ref{tab:model-summary}.
Each model corresponds to a particular instantiation of our framework, characterized by a combination of choices such as prediction task, codebook optimization and hard (point-mass) or soft assignments (soft-min) of $q$.
Specifically, we denote our variational predictive coding framework as Masked-VPC or Future-VPC, depends on 
the prediction task.
The framework thus allows us to compare different features in isolation.
Note that the original HuBERT model \citep{hsu2021hubert} consists of two or more training iterations.
Since we concerns training objective rather than the models, 
we use HuBERT Obj to indicate its training ``objective'' to avoid confusions.
The first few sets experiments compares the different losses
in the first iteration, and we leave the results of the second
iteration to last subsection.

\subsection{Comparing Optimization Methods}
\label{sec:opt}

We first focus on comparing the first two variants (HuBERT Obj, Masked-VPC) in Table~\ref{tab:model-summary}, which optimize the same loss.
As pointed out in Section \ref{sec:extensions}, softening the point mass, i.e., moving away from hard assignment in $k$-means to soft assignment, provides an opportunity to jointly optimize the cross entropy and the reconstruction.
Not only could we jointly optimize both terms, we also have the option to choose between exact marginalization and sampling.
When sampling, we only take a single sample and use Gumbel softmax to compute the gradient.
There are a total of three options: the original hard assignment in HuBERT Obj, soft assignment with exact marginalization, and soft assignment with sampling.
HuBERT uses $k$-means++ \citep{arthur2006k} as initialization, so we also compare random initialization and $k$-means++ initialization for the codebook.

\begin{figure}
  \centering
    \includegraphics[width=7cm]{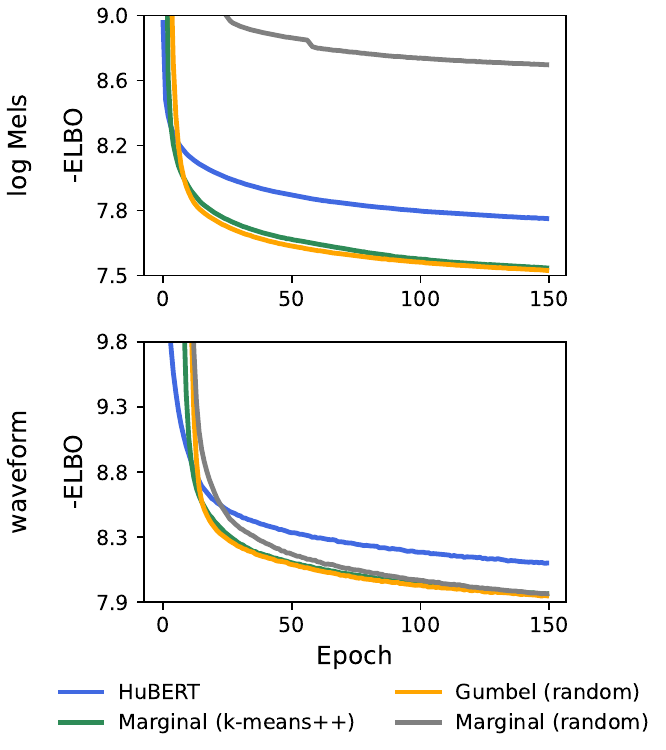}
    \caption{The training losses of different optimization approaches
      implemented by our own \textsc{Base} (top) and in fairseq \textsc{Base} (bottom).
      The final loss values can be found in Table~\ref{tab:ctc}.}
    \label{fig:optimization}
\end{figure}

Figure \ref{fig:optimization} (top) shows the variational objective (the negative ELBO in equation \ref{eq:vlb-mlm}) of training \textsc{Base} models on the 360-hour subset of Librispeech.
The HuBERT objective starts off low because of the offline $k$-means, but both exact marginalization and sampling improve over the HuBERT objective after training.
Similar to how HuBERT and BEST-RQ are sensitive to initialization \cite{chiu2022self}, marginalization ends up at drastically different final values depending on whether $k$-means++ is used.
Sampling (and taking the gradient with Gumbel softmax) turns out to be both efficient and fast converging.

To confirm the findings, we run the same experiments in fairseq\footnote{The experiments mostly follow
the default HuBERT configuration 
at \href{https://github.com/facebookresearch/fairseq/blob/main/examples/hubert/config/pretrain/hubert_base_librispeech.yaml}{\texttt{hubert\_base\_librispeech.yaml}},
using waveform as input.}
,
in which CNN is used to aggregate frame-wise features.
Figure \ref{fig:optimization} (bottom) shows the same trend, except that marginalization works out of the box with a randomly initialized codebook.
The final losses are presented in Table~\ref{tab:ctc}.
For the rest of the experiments, if not otherwise stated, we will only use Mel spectrograms as input.
We will also use sampling (and Gumbel softmax) to optimize our objective due to its faster convergence.

\subsection{Downstream Evaluation}

Given that we can improve pre-training, the next question is whether the improvement transfer to downstream tasks.
Besides, our framework allows for controlled experiments that only differ in loss functions while holding everything else, such as the model architecture and training hyperparameters, fixed.
Based on the connections to prior work in Section \ref{sec:related}, we will study two important design choices, i.e., comparing masked prediction to future prediction and comparing the contrastive loss in wav2vec 2.0 to the softmax in the HuBERT objective.

The utility of speech representations on downstream tasks are evaluated with probing \citep{oord2018representation, chung2019unsupervised, yang2022autoregressive, yang21c_interspeech}, where the pre-trained models are frozen and small classifiers are trained to complete tasks given the hidden vectors of one of the layers.

We conduct four probing tasks, phone classification, speaker verification, fundamental frequency tracking, and automatic speech recognition.
All settings follow \citet{yang2022autoregressive} unless otherwise noted.
For phone classification, we train a linear classifier to predict phone labels obtained from forced alignments on the Wall Street Journal \citep{paul1992design} (WSJ).
We report phone error rates (PERs) on \texttt{dev93} and \texttt{eval92}, using 
10\% of the training set \texttt{si284} for development.
For f0 tracking, we train a linear regression model to predict the f0 extracted with PYIN \citep{mauch2014pyin} on WSJ.
We report the root-mean-square error (RMSE) in Hz on \texttt{eval92}.
For speaker verification, we train a two-layer classifier to predict speaker identities on VoxCeleb1 \citep{nagrani2020voxceleb} and use the intermediate layer as the speaker embedding for speaker verification.
We report equal error rates (EERs) on the test set.
The reported numbers are with the best layer for each task and for each model.
More details are in Appendix \ref{appendix:downstream}.

\begin{table}
    \caption{Downstream probing results for different speech representations
        on phone classification, speaker verification, and f0 tracking.}
    \label{tab:downstream1}
    \centering
    \footnotesize
    \begin{tabularx}{0.98\linewidth}{lcccc}
            \toprule
                    &\multicolumn{2}{c}{PER $\downarrow$} &EER $\downarrow$ &RMSE $\downarrow$  \\
                    &dev93 &eval92 &voxceleb &eval92  \\[1pt]
            \midrule
            log Mel       &49.8          &50.0       &24.6        &38.4 \\
            i-vector      &-             &-          &15.7        &-    \\
            HuBERT Obj            &12.8          &12.6          &18.3      &23.4  \\
            Masked-VPC        &11.8          &11.3          &14.4      &20.9  \\
            \bottomrule
    \end{tabularx}
\end{table}

\paragraph{Better pre-training leads to better downstream performance}

We have shown that using soft assignment and sampling gives us the best pre-training loss.
To answer whether the improved pre-training transfer to downstream tasks,
we compare HuBERT Obj with the soft assignment variant optimized with sampling and Gumbel softmax (termed Masked-VPC).
Results on three tasks are shown in Table \ref{tab:downstream1}.
We see that improving pre-training indeed improves downstream tasks across the board.
ASR results are deferred to later sections, but the conclusion stands.

\begin{table}
    \caption{Downstream probing results comparing future prediction (Future-VPC) and masked prediction (Masked-VPC) on phone classification, speaker verification, and f0 tracking.}
    \label{tab:downstream2}
    \centering
    \footnotesize
    \begin{tabularx}{0.98\linewidth}{lcccc}
        \toprule
           &\multicolumn{2}{c}{PER $\downarrow$} &EER $\downarrow$ &RMSE $\downarrow$  \\
           &dev93 &eval92 &voxceleb &eval92  \\[1pt]
        \midrule                                                                                                                 
        Masked-VPC          &11.8          &11.3           &14.4     &20.9  \\
        Future-VPC          &16.0          &15.5           &13.6     &20.5  \\
        \bottomrule
    \end{tabularx}
\end{table}

\paragraph{Future prediction can be as strong as masked prediction on some downstream tasks}

As shown in Section \ref{sec:futurepred}, under our framework, it is possible (and relatively simple) to switch from masked prediction to future prediction while holding everything else fixed.
Future prediction is more relevant than masked prediction in certain scenarios, such as streaming or pre-training decoder-only Transformers.
\citet{misra21_interspeech}, for example, show that future prediction can be as strong as masked prediction for streaming ASR.
Results comparing future prediction (Future-VPC) and masked prediction (Masked-VPC) are shown in Table \ref{tab:downstream2}.
We find that, not surprisingly, future prediction is worse at phone classification compared to masked prediction.
However, future prediction is better than masked prediction on speaker verification and fundamental frequency tracking.

\begin{table}
    \caption{Downstream probing results comparing NCE (Masked-NCE), i.e., using a contrastive loss with the HuBERT approaches (HuBERT Obj and Masked-VPC) on phone classification, speaker verification, and f0 tracking.}
    \label{tab:downstream3}
    \centering
    \footnotesize
    \begin{tabularx}{0.98\linewidth}{lcccc}
    \toprule
       &\multicolumn{2}{c}{PER $\downarrow$} &EER $\downarrow$ &RMSE $\downarrow$  \\
       &dev93 &eval92 &voxceleb &eval92  \\[1pt]
    \midrule                                                                                                                 
    Masked-NCE       &12.2          &12.4      &20.2        &24.4       \\
    HuBERT Obj           &12.8          &12.6      &18.3        &23.4       \\
    Masked-VPC       &11.8          &11.3      &14.4        &20.9       \\
    \bottomrule
    \end{tabularx}
\end{table}

\paragraph{NCE is on par with HuBERT Obj}

To compare to the contrastive loss, we parameterize $q(z_i|x_i)$ and $p(z_i | x_{\setminus M})$ as detailed in Section \ref{sec:wav2vec}. 
We characterize the model in Table \ref{tab:model-summary} and denote it as Masked-NCE.
The entropy term is a constant because of the Dirac delta.
We follow \citet{baevski2020wav2vec} and take the negative samples for NCE from frames within the same batch.
The codebook is updated together with the contrastive loss with Gumbel softmax as in \citet{baevski2020wav2vec}.
Results are shown in Table \ref{tab:downstream3}.
NCE alone performs surprisingly well and closely matches HuBERT's results, while our approach is still superior.

\begin{table}[t]
  \caption{Probing results with lexicon-free seq2seq models on WSJ.
    No language models are used.}
  \label{tab:downstream_asr}
  \centering
  \begin{threeparttable}
  \footnotesize
  \addtolength{\tabcolsep}{-0.11em}
  \begin{tabularx}{0.94\linewidth}{lcccc}
    \toprule
      &\multicolumn{2}{c}{CER $\downarrow$} &\multicolumn{2}{c}{WER $\downarrow$}\\
            &dev93 &eval92  &dev93 &eval92\\[1pt]
    \midrule
    \multicolumn{4}{l}{\textbf{\textsc{Baseline}}}  \\[2pt]
    wav2letter++ \tnote{\dag}   &6.3   &4.1    &19.5     &13.9\\
    18L Transformer\tnote{\ddag}&- &- &22.2     &17.9\\
    \midrule                                                                                                                 
    \multicolumn{4}{l}{\textbf{\textsc{Probing with our seq2seq}}}  \\[2pt]
    log Mel     &6.8   &5.1  &18.2      &14.7 \\
    VQ-APC      &5.8  &5.2    &16.8      &14.9  \\
    Future-VPC   &5.4  &4.0    &15.5      &12.4  \\
    Masked-NCE   &5.0  &4.9    &14.5      &14.4  \\
    HuBERT Obj     &5.2  &5.0    &15.2      &14.5  \\
    Masked-VPC   &4.4  &3.6    &13.6      &11.4  \\
    \bottomrule
  \end{tabularx}
  \begin{tablenotes}\footnotesize
  \item[\dag] Numbers taken from \citet{baevski2019vq}.
  \item[\ddag] Numbers taken from \citet{higuchi2020mask}. 
  \end{tablenotes}
  \end{threeparttable}
\end{table}

\subsection{Automatic Speech Recognition}

Automatic speech recognition (ASR) has been the main driving force behind speech representation 
learning. In this section, we evaluate the quality of speech representations with two ASR settings, a probing setting with sequence-to-sequence (seq2seq) models and a fine-tuning setting with connectionist temporal classification (CTC).
We report CERs and WERs averaged over three runs. 

\paragraph{Probing with seq2seq}

In the probing setting, we follow \citet{yang2022autoregressive}, extracting the layer that achieves the best phone classification, and train a seq2seq model with character outputs on WSJ.\footnote{
More training details are in Appendix \ref{appendix:downstream}.}
Following the spirit of probing, we make sure that the improvement is solely from the learned representations 
and that the phonetic information is accessible, by not using a language model or a lexicon,
and by using a lightweight model.
Beam search is used with a beam size of~5.

Table~\ref{tab:downstream_asr} reports the ASR performance in terms of character error rates (CERs) and word error rates (WERs).
Despite being lightweight, our baseline model on log Mel features is by no means weak, and is on par with wav2letter++ in \citet{8683535} and 
better than a 18-layer Transformer baseline in \citet{higuchi2020mask}.
We then compare representations learned from the approaches characterized in Table \ref{tab:model-summary}, including two future prediction approaches (VQ-APC, Future-VPC) and three masked prediction approaches (Masked-NCE, HuBERT Obj, Masked-VPC).
We first find that representations learned from future prediction are not far behind those from masked prediction.
The second is that the improvement from HuBERT Obj to Masked-VPC (in Section \ref{sec:opt}) transfers well to ASR as well.

\begin{table}[t]
\centering
\captionof{table}{Results of ASR fine-tuned on WSJ using CTC
  for different optimization approaches during pre-training.
  No language models are used.
}
\label{tab:ctc}
\begin{threeparttable}
\footnotesize
\begin{tabularx}{0.94\linewidth}{lcccc}
  \toprule
     &$-$ELBO &dev93 &eval92\\[1pt]
  \midrule
  \multicolumn{3}{l}{\textbf{\textsc{Fine-Tuning (Ours)}}} &\\[1pt]
  HuBERT Obj               &7.79 &15.5 &12.5  \\
  Marginal ($k$-means++)     &7.50 &14.3 &11.0  \\
  Marginal (random)        &8.70 &15.0 &11.7  \\
  Gumbel (random)          &7.48 &14.1 &11.7  \\
  \midrule
  \multicolumn{3}{l}{\textbf{\textsc{Fine-Tuning (Fairseq)}}} &\\[1pt]
  HuBERT Obj               &8.14 &14.9 &11.8  \\
  Marginal ($k$-means++)     &7.91 &14.5 &11.1  \\
  Marginal (random)        &7.91 &14.2 &10.9  \\
  Gumbel (random)          &7.89 &14.3 &11.0  \\
  \bottomrule
\end{tabularx}
\end{threeparttable}
\end{table}

\paragraph{Fine-tuning with CTC}

To further confirm that improvement in pre-training transfers to ASR,
we fine-tune the pre-trained models on WSJ \citep{paul1992design} using CTC \citep{graves2006connectionist} with characters output. 
We fine-tune our \textsc{Base} for 100 epochs and fairseq \textsc{Base} for 200 epochs on \texttt{si284}.
Similar to \citet{lee2021intermediate,higuchi2020mask},
we average models from last 10 epochs after training to obtain final model parameters.
Table \ref{tab:ctc} summarizes the pre-training losses (-ELBO) reported in Figure~\ref{fig:optimization} and 
the corresponding WERs on \texttt{dev93} and \texttt{eval92}. The inference is done with greedy decoding.
The improvement in objective indeed transfers well to ASR in both settings.

To see whether the improvement is still present when decoding with a language model, Table~\ref{tab:ctc-lm} compared the results after using a 4-gram word language model \citep{heafield2013scalable}.
A beam size of 2000 is used for decoding with the 4-gram language model.
The improvement transfer well even with the use of a language model.

Lastly, to see the effect on small datasets, we follow \citet{baevski2020wav2vec} and evaluate phonetic recognition on TIMIT.
We observe the same trend in Table \ref{tab:ctc2} that our approach consistently performs better than HuBERT Obj.

\begin{table}[t]
\centering
\captionof{table}{Comparing ASR fine-tuned on WSJ using CTC with and without a 4-gram language model.
}
\label{tab:ctc-lm}
\begin{threeparttable}
\footnotesize
\begin{tabularx}{0.97\linewidth}{lccccc}
  \toprule
                                & \multicolumn{2}{c}{w/o LM} &\multicolumn{2}{c}{w/ LM} \\
                                & dev93 & eval92  &dev93 &eval92 \\
  \midrule
  \textbf{\textsc{Baseline}} &&&&\\[1pt]
    12L Transformer\tnote{\ddag} &20.1 &16.5 &- &-\\
    wav2letter++ \tnote{\dag}     &19.5     &8.57 &&\\
  \midrule
  \multicolumn{3}{l}{\textbf{\textsc{Fine-Tuning}}} &\\[1pt]
    Masked-NCE                     &14.7 &11.1  &7.8 &4.4 \\
    HuBERT Obj                     &15.5  &12.5 &7.3  &4.6\\
    Masked-VPC                     &14.1  &10.3 &6.8  &4.6\\
  \bottomrule
\end{tabularx}
\begin{tablenotes}\footnotesize
\item[\ddag] Numbers taken from \citet{lee2021intermediate}.
\item[\dag] Numbers taken from \citet{baevski2019vq}.
\end{tablenotes}
\end{threeparttable}
\end{table}

\begin{table}[t]
  \centering
  \caption{Phone recognition fine-tuned with CTC on TIMIT.}
  \label{tab:ctc2}
  \footnotesize
  \begin{tabular}{lccc}
    \toprule
                             &dev PER     &test PER $\downarrow$ \\
    \midrule
    \textbf{\textsc{Baseline}} &\\[1pt]
    12L Transformer &22.1 &24.1\\
    \midrule
    \multicolumn{2}{l}{\textbf{\textsc{Fine-Tuning}}} \\[1pt]
    Masked-NCE      &11.2  &12.9 \\
    HuBERT Obj      &10.2  &11.3 \\
    Masked-VPC      &9.5   &11.0 \\
    \bottomrule
  \end{tabular}
\end{table}

\subsection{Second-Iteration Training}

As detailed in Section \ref{sec:hubert-variants},
our framework can be extended to the second iteration
as in HuBERT \citep{hsu2021hubert}.
Following \citet{hsu2021hubert}, the Transformer encoder
from the previous iteration is held fixed in the second iteration.

We run the second iteration using fairseq, based on the its first iteration in Figure \ref{fig:optimization} (bottom). 
For the proposed approach, we choose Masked-VPC with $k$-means++ initialization from the first iteration, 
applying the same initialization to the second iteration training.
We adopt the 6-th layer of Transformers of HuBERT and Masked-VPC in the previous iteration, the layer that provides the best downstream ASR performance in Table \ref{tab:downstream_asr}.
We use a temperature $\tau$ of 10 to avoid $q$ being one-hot.
We train second-iteration HuBERT and Masked-VPC for 150 epochs, and evaluate them with CTC finetuning 
as in the previous section.

As shown in Table~\ref{tab:ctc-iter2}, the second iteration improves both approaches by a large margin.
In particular, we observe 4.1\% reduction of WER on \texttt{dev93} with our approach. 
Moreover, the improvements in objective transfers to the second iteration, in which 
Masked-VPC obtains 0.7\% and 0.3\% lower WERs than HuBERT on each set.
Similar trend is observed with an ngram language model for CTC decoding.

\begin{table}[t]
\centering
\captionof{table}{Second-iteration results of ASR fine-tuned on WSJ using CTC with or without an ngram LM.}
\label{tab:ctc-iter2}
\footnotesize
\begin{tabular}{lccccc}
  \toprule
  & \multicolumn{2}{c}{w/o LM} & \multicolumn{2}{c}{w/ LM}\\
  &dev93 &eval92 &dev93 &eval92\\[1pt]
  \midrule
  \multicolumn{3}{l}{\textbf{\textsc{First iteration}}} &\\[1pt]
  HuBERT Obj                  &14.9 &11.8 &7.9 &6.0\\
  Masked-VPC                &14.5 &11.1 &7.8 &5.3 \\
  \midrule
  \multicolumn{3}{l}{\textbf{\textsc{Second iteration}}} &\\[1pt]
  HuBERT Obj                &11.1 &8.2  &6.2 &4.2 \\
  Masked-VPC                &10.4 &7.9  &5.9 &3.8 \\
  \bottomrule
\end{tabular}
\end{table}

\section{Conclusion}

In this work, we provide an underlying principle for the HuBERT objective---a
variational view of predictive coding.
We show the utility of this framework by identifying opportunities to improve the HuBERT objective, i.e., its parameterization and optimization.
Evaluating across several downstream tasks using probing and fine-tuning, we empirically show that the improved pre-training learns better speech representations.
We further show that the predictive coding framework is general and has many connections to existing self-supervised objectives.
Predictive coding has fruitful and mature theoretical underpinnings, while the theory of self-supervised learning is still in its infancy.
We hope the making the connections among self-supervised learning and predictive coding clear as is done in this work helps advance the understanding of both.

%\bibliography{tacl2021}
%\bibliographystyle{acl_natbib}

\balance{
\bibliographystyle{acl_natbib}
\bibliography{tacl2021}
}
\newpage
\nobalance
\appendix
\begin{table*}[ht]
\centering
\small
\begin{tabular}{lccccc}
\toprule
 & VQ-APC & Future-VPC & Masked-NCE & HuBERT & Masked-VPC\\
\midrule
PC/ASR & 8      & 8          & 9           & 8      & 8       \\
f0 & 3      & 3          & 2           & 2      & 3       \\
SV & 4      & 3          & 3           & 3      & 3       \\
\bottomrule
\end{tabular}
\caption{Layers selected for downstream experiments for each model variant.}
\label{tab:layers}
\end{table*}

\section{Appendix}

\subsection{Negative Free Energy of Predictive Coding}
\label{appendix:proof}

The proof of equation \ref{eq:vlb} mirrors those in variational autoencoders \citep{Kingma2014}.
We start with the KL divergence between the posterior distribution $p(z|x_A, x_B)$
and the auxiliary distribution $q(z|x_B)$,
\begin{align}
&\KL\big[q(z|x_B)\|p(z|x_A,x_B)\big] \\
&\quad = \mathbb{E}_{z \sim q} \left[\log \frac{q(z | x_B)}{p(x_B | z) p(z | x_A)} \right] + \log p(x_B|x_A) \notag
\label{eq:proof}
\end{align}
where we assume $x_B \ind x_A \mid z$ or $p(x_B|x_A, z) = p(x_B|z)$.
Because the KL divergence is always positive, we obtain
\begin{align}
&-\log p(x_B | x_A) \\
&\quad \leq \KL \big[q(z|x_B)\|p(z|x_A)\big] - \mathbb{E}_{z \sim q}[\log p(x_B | z)] \notag
\end{align}

\subsection{Future Prediction}

The proof of equation \ref{eq:vlb-ar} involves unrolling $p(x_B, z | x_A)$ or $p(x_{\geq t}, z_{\geq t} | x_{< t})$, where $x_B = x_{\geq t}$ and $x_A = x_{< t}$.
Based on the definition of conditional probability, we have
\begin{align}
&p(x_{\geq t}, z_{\geq t} | x_{< t}) = \prod_{i=t}^T p(x_i, z_i | x_{< i}, z_{t:i-1}) \\
&\quad = \prod_{i=t}^T p(x_i | x_{< i}, z_{t:i}) p(z_i | x_{< i}, z_{t:i-1}) \\
&\quad = \prod_{i=t}^T p(x_i | z_i) p(z_i | x_{< i}),
\end{align}
where the last line makes two reasonable assumptions, $x_i \ind z_{t:i-1} \mid z_i$ and $z_i \ind z_{t:i-1} \mid x_{< i}$.
The variable $z_i$ should represent $x_i$ well without relying on previous latent variables $z_{t:i-1}$.
Given the history $x_{< i}$, computing the representation $z_i$ should not depend on the previous latent variables $z_{t:i-1}$.

\subsection{Pre-Training Recipes}
\label{appendix:recipe}

We set a maximum length of 1400 frames per utterance, 
corresponding to about 28 seconds. The learning rate is fixed to $10^{-4}$ under the Adam optimizer,
no learning rate schedule is applied.
We pre-train
\textsc{Base} models on a single A40 with a batch size of 16 for 150 epochs. 
We set model dimension to 768, with inner dimension of feedforward netowrks being 3072.
A dropout probability of 0.1 is applied.

The temperature is set to 1 without annealing for variational training.
In wav2vec 2.0, the similarity value is re-scaled by dividing it with 0.1, 
the temperature for Gumbel-Softmax \citep{jang2016categorical} is annealed from 2 to a minimum of 0.5 by a decay rate of 0.999995,
i.e., the $\tau$ in \Eqref{eq:softmax-q}.
We use 100 negative samples following \citet{baevski2020wav2vec}.

There are a few architectural differences between our setting and \citet{baevski2020wav2vec}.
We employ a single codebook in our wav2vec 2.0 for quantization.
Rather than Post-LN Transformers, 
we use Pre-LN Transformers for pre-training
and remove the warm-up stage \citep{geiping2023cramming,xiong2020layer}.
We use sinusoidal positional embeddings rather than relative position embeddings used in 
\citet{baevski2020wav2vec}.

\subsection{Downstream Evaluation}
\label{appendix:downstream}

We provide additional details on the experimental setups of downstream tasks. 
The layers chosen for each task is noted in Table~\ref{tab:layers}.

\paragraph{Phone Classification (PC)}
We freeze the pre-trained model and only train a linear layer with a learning rate of $10^{-3}$ for 10 epochs.

\paragraph{Speaker Verification (SV)}

We average frame representations to obtain utterance-level representations, and simply
employ two linear layers to predict 1251 speakers following \citep{Fan2020ExploringW2}. 
The linear classifier is optimized with a learning rate of $10^{-3}$ for 10 epochs.
After training, we take the output of the first linear layer as speaker vectors for speaker verification.
We set the dimension of speaker vectors to 512.

\paragraph{F0 Tracking (F0)}
We train a linear regression layer with a learning rate of $10^{-3}$ for 10 epochs.
We set the minimum and maximum frequency to 50 Hz and 600 Hz respectively.

\paragraph{Automatic Speech Recognition (ASR)}

We use a lightweight sequence-to-sequence encoder for downstream ASR.
The encoder contains two convolutional layers with $(32, 32)$ channels and $(2, 1)$ strides,
and a 4-layer, 256-dimensional bidirectional GRU. The decoder is a unidirectional 256-dimensional GRU.

We adopt a fixed scheduled sampling probability of 0.4 during training. 
We use Adam with learning rates of $10^{-4}$ for all s2s models. 
We employ a dropout rate of 0.2, and a label smoothing rate of 0.1 for regularization. 
We train seq2seq models for 100 epochs, 
and lower the learning rate with a factor of 0.1 for another 20 epochs.

\end{document}